\documentclass[9pt,journal]{IEEEtran}
\ifCLASSINFOpdf
\else
   \usepackage[dvips]{graphicx}
\fi
\usepackage{amsmath}
\newtheorem{theorem}{Theorem}

\newtheorem{lemma}{Lemma}
\newtheorem{definition}{Definition}
\newtheorem{assumption}{Assumption}
\newtheorem{corollary}{Corollary}

\begin{document}
%
% paper title
% Titles are generally capitalized except for words such as a, an, and, as,
% at, but, by, for, in, nor, of, on, or, the, to and up, which are usually
% not capitalized unless they are the first or last word of the title.
% Linebreaks \\ can be used within to get better formatting as desired.
% Do not put math or special symbols in the title.
\title{Equivalent Stability Notions, Lyapunov Inequality, and\\ Its Application in Discrete-Time Linear Systems with
Stochastic Dynamics Determined by an i.i.d.~Process}
%
%
% author names and IEEE memberships
% note positions of commas and nonbreaking spaces ( ~ ) LaTeX will not break
% a structure at a ~ so this keeps an author's name from being broken across
% two lines.
% use \thanks{} to gain access to the first footnote area
% a separate \thanks must be used for each paragraph as LaTeX2e's \thanks
% was not built to handle multiple paragraphs
%

\author{Yohei~Hosoe %~\IEEEmembership{Member,~IEEE,}
        and~Tomomichi~Hagiwara %,~\IEEEmembership{Senior~Member,~IEEE,}
%        and~Dimitri~Peaucelle %,~\IEEEmembership{Member,~IEEE}% <-this % stops a space
%
\thanks{This work was
supported by JSPS KAKENHI Grant Number 17K14700.}
\thanks{Y.~Hosoe and T.~Hagiwara are with the Department of Electrical Engineering, 
        Kyoto University, Nishikyo-ku, Kyoto 615-8510, Japan
        (e-mail: hosoe@kuee.kyoto-u.ac.jp).}% <-this % stops a space
%\thanks{D.~Peaucelle is with LAAS-CNRS, Universit\'{e} de Toulouse,
%CNRS, Toulouse, France.}% <-this % stops a space
%\thanks{Manuscript received April 19, 2005; revised August 26, 2015.}}
}

\maketitle

% As a general rule, do not put math, special symbols or citations
% in the abstract or keywords.
\begin{abstract}
This paper is concerned with stability analysis and synthesis for
discrete-time linear systems with stochastic dynamics.
Equivalence is
first proved for three stability notions under
some key assumptions on the randomness behind the systems.
In particular, we use the assumption that the stochastic process
determining the system dynamics is independent and identically distributed (i.i.d.)
with respect to the discrete time.
Then, a Lyapunov inequality condition is derived for 
stability in a
necessary and sufficient sense.
 Although our Lyapunov inequality will involve decision variables contained in the
expectation operation, an idea is provided to solve it as a standard
linear matrix inequality; the idea also plays an important role in state feedback synthesis based on the
 Lyapunov inequality.
Motivating numerical examples are
further discussed as an application of our approach.
\end{abstract}

% Note that keywords are not normally used for peerreview papers.

% For peer review papers, you can put extra information on the cover
% page as needed:
% \ifCLASSOPTIONpeerreview
% \begin{center} \bfseries EDICS Category: 3-BBND \end{center}
% \fi
%
% For peerreview papers, this IEEEtran command inserts a page break and
% creates the second title. It will be ignored for other modes.
\IEEEpeerreviewmaketitle

%%%%%%%%%%%%%%%%%%%%%%%%%%%%%%%%%%%%%%%%%%%%%%%%%%%%%%%%%%%%%%%%%%%%%%%%%%%%%%%%
\section{Introduction}

Recent advances in computer power are leading to
demands for extending the frontier of
control technologies to cover wider classes of systems.
Stimulated by such a trend, this paper focuses on
the class of discrete-time linear systems whose state transition is determined
only randomly,
also called discrete-time
linear random dynamical systems in the field of analytical dynamics \cite{Arnold-book}.
Randomness is fairly common in various kinds of phenomena
(e.g., packet interarrival times in networked
systems \cite{Paxson-IEEETN95} and failure occurrences in distributed systems \cite{Finkelstein-book}),
and discarding the information about it in modeling
might lead to the situation where the controllers designed with
the resulting models do not achieve the expected
performance for the original objectives.
Hence, if the randomness behind the real objects is essential, it
should also be modeled and exploited in controller synthesis.

Behaviors and properties of random dynamical systems have been
extensively studied, e.g., in \cite{Yu-PRL90,Arnold-DSS98,Wang-JDE12}.
However, studies dealing with such systems in control
problems are still rare.
Our ultimate goal is to develop a versatile practical framework for
controlling such systems by restricting our attention to the
linear case.
In particular, we aim at developing a systematic analysis
and synthesis approach based on linear matrix inequality (LMI) conditions,
as in the existing studies on deterministic linear systems \cite{Boyd-book,Scherer-TAC97,Svariable-Ebihara-book}.
As a step toward such a goal, this paper
first shows the equivalence of some stability notions and
derives the Lyapunov inequality condition for 
stability of discrete-time linear random dynamical systems under
some key assumptions.
Our Lyapunov inequality characterizes stability of
such systems in a necessary and sufficient sense
and will be a basis for further advanced analysis and synthesis using LMIs.

The state transition of our discrete-time linear random dynamical
systems can
be seen as
determined by an underlying stochastic process, and we assume in this paper
that the process is independent and identically distributed (i.i.d.)
with respect to the discrete time (hence the process naturally becomes
stationary and ergodic \cite{Knill-book}); this assumption will play
a key role in showing the above stability equivalence.
This system class contains various kinds of linear
stochastic systems studied in the literature.
For example, systems with state-multiplicative noise \cite{Boyd-book,Gershon-Auto08}
and switched systems \cite{Geromel-IJC06} with i.i.d.\ switching signals (which
correspond to Markov jump systems \cite{Costa-book} with transition probability uniform
in all current modes) belong to this class.
Hence, the results in this paper can be seen as a generalization of
those for such particular systems,
and would become a sort of center point for bridging and unifying 
the associated existing results.

For reference, we briefly summarize the technical aspect of the contributions
in this paper through the comparison with the closely related earlier
studies \cite{Costa-TAC14,Hosoe-TAC18}.
In \cite{Costa-TAC14}, a necessary and sufficient stability condition is
shown for discrete-time linear systems with stochastic dynamics
determined by a stationary Markov process.
Since i.i.d.~processes are a special case of stationary Markov
processes, one might consider that our results could be covered by those for
Markov jump systems.
However, this is not true because the above earlier results are derived
with the assumption that the maximum
singular value of the random coefficient matrix
(depending on the Markov process)
is essentially bounded,
which makes it impossible to deal with random coefficient matrices
involving, e.g., normally distributed elements.
This paper will use a milder assumption for this part (see
Assumption~\ref{as:bound} introduced later).
Hence, our results cannot generally be covered by those in
\cite{Costa-TAC14} (and vice versa).
In the earlier study \cite{Hosoe-TAC18} of the authors, a sufficient
stability condition was shown
as a part of the contributions for essentially the same
stochastic systems in the present paper.
However, only exponential stability was dealt with and the necessity
assertion of the condition was not discussed even for 
that stability notion.
This paper will complement this earlier study
from several viewpoints.

The contents of this paper are as follows.
In Section~\ref{sc:sys}, the stochastic system to be dealt with in
this paper is described, and three stability notions are
introduced:
asymptotic stability, exponential stability \cite{Kozin-Auto69} and quadratic stability.
Then, the equivalence of those stability notions
is proved in Section~\ref{sc:equiv}.
Then, the Lyapunov inequality is derived in Section~\ref{sc:lyap}
as a necessary and sufficient condition for quadratic stability.
Since our Lyapunov
inequality will involve decision variables contained in the operation of
expectation, we will also provide ideas for solving the inequality as a
standard LMI involving no expectation operation.
The stabilization feedback synthesis based on the Lyapunov inequality is 
further discussed in Section~\ref{sc:syn}.
Finally, two numerical examples are provided for our
stability analysis and synthesis in Section~\ref{sc:example}.
The first example has the role of demonstrating that the Lyapunov
inequality indeed gives a necessary and sufficient condition
not only for quadratic stability but for exponential stability, as is
theoretically indicated.
On the other hand, the second example is provided for showing the
potential of our approach by tackling a challenging problem;
more specifically,
we consider stabilizing the discrete-time system obtained through
discretizing a continuous-time deterministic linear system with a
randomly time-varying sampling interval, which is inspired by the studies
on aperiodic sampling \cite{Montestruque-TAC04,Hetel-Auto17} (related,
e.g., to packet interarrival times in networked systems).

We use the following notation in this paper.
${\bf R}$, ${\bf R}_+$ and ${\bf N}_0$ denote the set of real numbers,
that of positive real numbers
and that of non-negative integers, respectively.
${\bf R}^n$ and ${\bf R}^{m\times n}$ denote the set of $n$-dimensional real
column vectors and that of $m\times n$ real matrices, respectively.
${\bf S}^{n\times n}$ and ${\bf S}^{n\times n}_{+}$ denote 
the set of $n\times n$ symmetric matrices and that of 
$n\times n$ positive definite matrices, respectively.
$\sigma_{\rm max}(\cdot)$ and $\sigma_{\rm min}(\cdot)$
denote the maximum and minimum singular values of the matrix $(\cdot)$, respectively.
$||(\cdot)||$ denotes the Euclidean norm of
the vector $(\cdot)$.
${\rm row}(\cdot)$ denotes the vectorization of the matrix $(\cdot)$ in the row
direction, i.e., ${\rm row}(\cdot)=[{\rm row}_1(\cdot),\ldots,{\rm
row}_m(\cdot)]$ where $m$ is the number of rows of the matrix and ${\rm row}_i(\cdot)$ denotes the $i$th row.
$\otimes$ denotes the Kronecker product.
${\rm diag}(\cdot)$ denotes the (block-)diagonal matrix.
$E[(\cdot)]$ 
denotes the expectation of the random variable $(\cdot)$; this notation
is also used for the expectation of the random matrix $(\cdot)$.
If $s$ is a random variable obeying the distribution ${D}$,
then we represent it as $s \sim {D}$.

\section{Stability of Discrete-Time Linear Systems with Stochastic
 Dynamics}
\label{sc:sys}

\subsection{Discrete-Time Linear Systems with Stochastic
 Dynamics}

Let us consider the $Z$-dimensional discrete-time stochastic process
$\xi$, which is the sequence of $Z$-dimensional
random vectors $\xi_k$ with respect to the discrete time $k\in {\bf
N}_0$, and make the following key assumption on it.

\begin{assumption}
\label{as:iid}
$\xi_k$ is independent and identically distributed (i.i.d.) with
 respect to $k\in {\bf N}_0$.
\end{assumption}

This assumption naturally makes $\xi$ stationary and ergodic \cite{Knill-book}.
For this stochastic process $\xi$, we denote
the cumulative distribution
 function of $\xi_k$ and the corresponding support by 
${\cal F}(\xi_k)$ and ${\boldsymbol {\mit\Xi}}$, respectively.
By definition, ${\boldsymbol {\mit\Xi}} \subset {\bf R}^Z$, 
and	${\boldsymbol {\mit\Xi}}$ corresponds to the set of values
that $\xi_k$ can take.

Let	us further consider the discrete-time linear system
\begin{equation}
x_{k+1} = A(\xi_k) x_k,
\label{eq:fr-sys}
\end{equation}
where $x_k \in {\bf R}^n$, $A:{\boldsymbol {\mit\Xi}} \rightarrow
{\bf R}^{n\times n}$, and the initial state $x_0$ is assumed to be deterministic.
Since $A(\xi_k)$ is a random matrix (while $A(\cdot)$ itself is a
deterministic mapping), the dynamics
of the above system is stochastic.
To ensure mathematical rigor throughout this paper,
we make the following assumption on 
the coefficient matrix $A(\xi_k)$ of the system.

\begin{assumption}
\label{as:bound}
The squares of elements of 
$A(\xi_k)$ are all Lebesgue integrable, i.e.,
\begin{align}
&
E[A_{ij}(\xi_k)^2]<\infty\ \ (\forall i, j = 1,\ldots,n),
\label{eq:as-bound}
\end{align}
where $A_{ij}(\xi_k)$ denotes the $(i,j)$-entry of $A(\xi_k)$.
\end{assumption}

In this paper, we say that the expectation of a random variable is
well-defined, if the random variable is Lebesgue integrable; hence,
$E[A_{ij}(\xi_k)^2]$ satisfying (\ref{eq:as-bound}) is said to be
well-defined.
This term is also used for the expectation of a random matrix
when its elements are all Lebesgue integrable.

The aim of this paper is to develop a theoretical basis of stability analysis
and synthesis for system (\ref{eq:fr-sys}) with $\xi$ satisfying
Assumptions~\ref{as:iid} and \ref{as:bound}.
Since we have introduced no essential restrictions on ${\cal F}(\cdot)$ and
$A(\cdot)$, this system covers a wide class of
discrete-time linear systems with stochastic dynamics; 
indeed, system (\ref{eq:fr-sys}) is the most general for representing the discrete-time
linear finite-dimensional systems with stochastic dynamics (without
additive inputs) under Assumptions~\ref{as:iid} and \ref{as:bound}.
Assumption~\ref{as:bound} would not become a problem from the practical
viewpoint, and hence,
the only essential restriction on the system is Assumption~\ref{as:iid}, which plays a
crucial role throughout this paper.

\subsection{Stability Notions}
\label{ssc:stab}
We next introduce three stability notions for
system (\ref{eq:fr-sys}) with $\xi$ satisfying
Assumptions~\ref{as:iid} and \ref{as:bound}.
The first and second notions are asymptotic stability and exponential
stability \cite{Kozin-Auto69} defined as follows.

\begin{definition}[Asymptotic Stability]
\label{df:asym}
The system (\ref{eq:fr-sys}) with $\xi$ satisfying
Assumptions~\ref{as:iid} and \ref{as:bound} is said to be stable in the second moment
 if for each positive $\epsilon$, there exists
 $\delta=\delta(\epsilon)$ such that
\begin{align}
&
\|x_0\|^2 \leq \delta(\epsilon) \Rightarrow
E[\|x_k\|^2] \leq \epsilon\ \ (\forall k
 \in {\bf N}_0).\label{eq:stab-def}
\end{align}
In addition, the 
system is said to be
%globally uniformly
 asymptotically stable
 in the second moment if the system is stable in the second moment and
\begin{align}
&
E[\|x_k\|^2] \rightarrow 0\ \ {\rm as}\ \ k\rightarrow \infty\ \ (\forall x_0 \in {\bf R}^n).\label{eq:asy-def}
\end{align}
\end{definition}

\begin{definition}[Exponential Stability]
\label{df:expo}
The system (\ref{eq:fr-sys}) with $\xi$ satisfying
Assumptions~\ref{as:iid} and \ref{as:bound} is said to be exponentially stable
 in the second moment if there exist $a\in {\bf R}_+$ and $\lambda \in
 (0,1)$ such that
\begin{align}
&
\sqrt{E[||x_k||^2]} \leq a ||x_0|| \lambda^k\ \ \ (\forall k \in {\bf
N}_0, \forall x_0 \in {\bf R}^n).
\label{eq:exp-def}
\end{align}
\end{definition}

The second-moment asymptotic (resp.\ exponential) stability defined above is also
called asymptotic (resp.\ exponential) mean square stability \cite{Kozin-Auto69},
 and is widely used in the field of stochastic systems control.
In Definition~\ref{df:expo},
$\lambda$ is an upper bound of the convergence rate
with respect to
the sequence 
$\left(\sqrt{E[||x_k||^2]}\right)_{k \in {\bf N}_0}$.

Compared to the above two notions, the following third notion might not be
major in the field of stochastic systems control but is closely
related to our main arguments.

\begin{definition}[Quadratic Stability]
\label{df:quad}
The system (\ref{eq:fr-sys}) with $\xi$ satisfying
Assumptions~\ref{as:iid} and \ref{as:bound} is said to be quadratically stable
if there exist $P\in {\bf S}^{n\times n}_+$ and $\lambda\in (0,1)$ such
 that
\begin{align}
&
E[x_{k+1}^T P x_{k+1}]\leq \lambda^2 E[x_k^T P x_k]\ \ \ (\forall k \in {\bf
N}_0, \forall x_0 \in {\bf R}^n).
\label{eq:quad-def}
\end{align}
\end{definition}

In the above definition of quadratic stability, $V(x_k)=E[x_k^T P x_k]$ is the quadratic Lyapunov function
described with the Lyapunov matrix $P$,
and (\ref{eq:quad-def}) requires the existence of a Lyapunov function (i.e.,
$P$) that decays no slower than
the rate $\lambda^2\ (<1)$, as is the
case with deterministic systems \cite{Boyd-book}.

Here, to ensure mathematical rigor, we show that 
Assumptions~\ref{as:iid} and \ref{as:bound} lead to
the well-definedness of the expectations referred to in the above
definitions.
As a step for this end, we first note two facts.
The first fact is that
if $s_1\leq s_2$ (resp.\ $s_1< s_2$) for each
sample of the pair of the two random variables $s_1$ and $s_2$,
then $E[s_1]\leq E[s_2]$ (resp.\ $E[s_1]< E[s_2]$);
since this fact is almost trivial, we use it throughout this paper
without any specific notes.
Compared to the first fact, the second fact might not be trivial and we
would like to summarize it as in the following lemma, 
which can be shown with the Cauchy-Schwarz inequality.

\begin{lemma}
\label{lm:product-bound}
If the expectations of $s_1^2$ and
 $s_2^2$ are well-defined for the random variables $s_1$ and $s_2$, then the expectation of $s_1 s_2$ also is.
\end{lemma}

Then, by using the above two facts, 
we can obtain the following result:
for the random vector $s_1$ and the square random matrix $S_2$ (of the
compatible size) such that 
$s_1$ and $S_2$ are independent of each other and $E[S_2]$ is
well-defined, the expectation $E[s_1^T S_2 s_1]\ (=E[s_1^T E[S_2] s_1])$ is well-defined
if $E[\|s_1\|^2]$ is.
Hence, by taking $s_1=x_k$ and $S_2=A(\xi_k)^T A(\xi_k)$
under Assumptions~\ref{as:iid} and \ref{as:bound},
we can show that
if $E[\|x_k\|^2]$ is well-defined, then $E[\|x_{k+1}\|^2]$ also is;
the well-definedness of $E[S_2]=E[A(\xi_k)^T A(\xi_k)]$ can be 
ensured by Lemma~\ref{lm:product-bound} under Assumption~\ref{as:bound}.
A recursive use of this result leads to the well-definedness of
$E[\|x_k\|^2]$ for every $k\in {\bf N}_0$.
The well-definedness of $E[x_k^T P x_k]$ can be ensured in a similar fashion.
Hence, the expectations in Definitions~\ref{df:asym} through
\ref{df:quad} are all well-defined under Assumptions~\ref{as:iid} and
\ref{as:bound}.

\section{Equivalence of Three Stability Notions}
\label{sc:equiv}

Three stability notions were introduced in the preceding section:
asymptotic stability, exponential stability and quadratic stability.
Since quadratic stability is usually introduced as a notion
related with   
deterministic time-invariant Lyapunov matrices
(as in Definition~\ref{df:quad}),
it is not equivalent to asymptotic stability and
exponential stability, in general.
For example, in the case of deterministic linear time-varying systems,
such equivalence is known to fail \cite{Daafouz-SCL01}.
Hence, one might be concerned about the possibility of a similar
situation for the system (\ref{eq:fr-sys}) since it can be viewed
as a deterministic linear time-varying system when we discard
the information about
underlying randomness.
However, we can actually 
establish equivalence of
all these notions also for the stochastic system (\ref{eq:fr-sys})
(and the present stability definitions), {\it
provided that Assumptions~\ref{as:iid} and \ref{as:bound} are satisfied}, as in the
case with deterministic linear time-invariant (LTI) systems.
Showing this non-trivial equivalence is one of the main
results in this paper.

\subsection{Equivalence between Asymptotic Stability and Exponential Stability}

We first give the proof
of the following theorem
about equivalence between asymptotic stability and exponential stability
(similar equivalence is known to hold for deterministic linear
systems \cite{Vidyasagar-book}).

\begin{theorem}
\label{th:eqv-asym-expo}
Suppose $\xi$ satisfies Assumption~\ref{as:iid} and $A(\xi_k)$ satisfies
 Assumption~\ref{as:bound}.
The following two conditions are equivalent.
\begin{enumerate}
 \item 
The system (\ref{eq:fr-sys}) is asymptotically stable in the second moment.
\item
The system (\ref{eq:fr-sys}) is exponentially stable in the second moment.
\end{enumerate}
\end{theorem}

\begin{IEEEproof}
2$\Rightarrow$1:
It follows from (\ref{eq:exp-def}) and $0<\lambda<1$ that
\begin{align}
&
E[\|x_k\|^2]\leq a^2 \|x_0\|^2\ \ (\forall k \in {\bf
N}_0).
\label{eq:proof-th1-a2x2}
\end{align}
This leads us to (\ref{eq:stab-def}) with
$\delta(\epsilon)=\epsilon/a^2$, which
means the second-moment stability of the system.
In addition, (\ref{eq:asy-def}) readily follows from
 (\ref{eq:exp-def}) since $0<\lambda<1$.
Hence, by definition, the system is asymptotically stable in the second moment.

\medskip
1$\Rightarrow$2:
Linearity of the system (\ref{eq:fr-sys}) frequently used in
this part of the proof is not explicitly referred to so as not to make the arguments verbose.
We first introduce the decomposition
\begin{align}
&
x_0=\beta \sum_{i=1}^n a_i \sigma_i e^{(i)}
\end{align}
with the scalars $\beta$, 
$a_i\geq 0\ (i=1,\ldots,n)$ satisfying $\sum_{i=1}^n a_i=1$,
the integers $\sigma_i\in \{-1,1\}\ (i=1,\ldots,n)$
and the standard basis vectors $e^{(i)}\ (i=1,\ldots,n)$ for the
$n$-dimensional Euclidean space.
By definition, we have
\begin{align}
\|x_0\|^2
%&=
%\beta^2 \left\|\sum_{i=1}^n a_i \sigma_i e^{(i)}\right\|^2 \notag \\
&=
\beta^2(a_1^2+\ldots +a_n^2)
\geq
\beta^2/n.\label{eq:th1-pf-x0low}
\end{align}
%
%it follows that  $\beta^2\leq n$ if $\|x_0\|^2\leq 1$.
Associated with this decomposition of $x_0$, we can also decompose the corresponding
$x_k$ as
\begin{align}
&
x_k=\beta \sum_{i=1}^n a_i \sigma_i x^{(i)}_k,
\end{align}
where $x^{(i)}_k$ is the state at $k$ for the initial state
$x_0=e^{(i)}$.
It follows from (\ref{eq:asy-def}) that
there exists $K\in {\bf N}_0$ such that
\begin{align}
&
E[\|x^{(i)}_k\|^2]\leq 1/(2n^2)\ \ (i=1,\ldots,n; \forall k\geq K).
\end{align}
Then, we have
\begin{align}
E[\|x_k\|^2]
&=
\beta^2
E\left[
\left\|\sum_{i=1}^n a_i \sigma_i x^{(i)}_k\right\|^2\right] \notag \\
&\leq
\beta^2 
E\left[
\sum_{i=1}^n a_i \|\sigma_i x^{(i)}_k\|^2\right] \notag \\
&=
\beta^2 \sum_{i=1}^n a_i E[\|x^{(i)}_k\|^2] \notag \\
&\leq
\beta^2/(2n)\ \ (\forall k \geq K),\label{eq:th1-pf-xkup}
\end{align}
where the first inequality follows from Jensen's inequality.
Hence, it follows from (\ref{eq:th1-pf-x0low}) and (\ref{eq:th1-pf-xkup})
that
%%
%\begin{align}
%\|x_0\|^2\leq 1 &\Rightarrow \beta^2\leq n \notag \\
%&\Rightarrow E[\|x_k\|^2]\leq1/2\ \ (\forall k\geq K)
%\label{eq:proof-th1-112}
%\end{align}
%%
%for the same $K$,
%which implies
%For each non-zero $x_0 \in {\bf R}^n$, let us define
% $\widetilde{x}_0:=\|x_0\|^{-1}x_0$ as the new initial state of system (\ref{eq:fr-sys}), which satisfies
% $\|\widetilde{x}_0\|\leq 1$.
%Then, by (\ref{eq:proof-th1-112}), we see that
%the corresponding state $\widetilde{x}_k$ satisfies
% $E[\|\widetilde{x}_k\|^2]\leq1/2\ (\forall k\geq K)$.
%Since the system is linear, this implies $E[\|x_k\|^2]\leq\|x_0\|^2/2\
% (\forall k\geq K)$, that is,
%
\begin{align}
&
E[\|x_K\|^2]\leq\|x_0\|^2/2 \label{eq:proof-th1-rec-orig}
\end{align}
%
%for the original initial state $x_0$.
for the same $K$.
Since this inequality holds regardless of $x_0 \in {\bf R}^n$
and since $\xi$ satisfies Assumption~\ref{as:iid} (in particular,
the stationarity assumption of $\xi$), 
we further have
\begin{align}
&
E[\|x_{k+K}\|^2]\leq E[\|x_k\|^2]/2\ \ (\forall k\in {\bf N}_0).
\label{eq:proof-th1-rec}
\end{align}
For each $k\in {\bf N}_0$, take $j$ and $c$ such that 
$k=c+jK\ (0\leq c<K)$.
Then, a recursive use of (\ref{eq:proof-th1-rec}) leads to
\begin{align}
E[\|x_k\|^2]&=E[\|x_{c+jK}\|^2] \notag\\
&\leq
E[\|x_{c}\|^2]/2^j
\notag \\
&=
2^{c/K}E[\|x_{c}\|^2](2^{-1/K})^k
\notag \\
&\leq 2E[\|x_{c}\|^2](2^{-1/K})^k \ \ (\forall k
 \in {\bf N}_0).
\label{eq:proof-th1-expo}
\end{align}
Since Assumptions~\ref{as:iid} and \ref{as:bound} ensure that
$E[\|x_{c}\|^2]$ is well-defined for every $c \in [0,K)$ (from the
 arguments in the preceding subsection),
there exists a bounded positive scalar $\alpha_K$
 such that
\begin{align}
&
E[\|x_{c}\|^2] \leq \alpha_K \|x_0\|^2 \ \ (\forall c \in [0,K)).
\end{align}
This, together with (\ref{eq:proof-th1-expo}), leads us to
\begin{align}
E[\|x_k\|^2]
&\leq 2\alpha_K \|x_{0}\|^2(2^{-1/K})^k \ \ (\forall k
 \in {\bf N}_0),
\label{eq:proof-th1-expo-fin}
\end{align}
which implies the 
existence of $a=2\alpha_K$ and $\lambda=2^{-1/K}$ such
 that $a>0$, $0<\lambda<1$ and (\ref{eq:exp-def}) hold.
Hence, by definition, the system is exponentially stable in the second moment.
This completes the proof.
%Here, from (\ref{eq:stab-def}) with $\epsilon=1$,
%there exists $\delta$ such that
%%
%%
%\begin{align}
%&
%\|x_0\|^2 \leq \delta \Rightarrow
%E[\|x_k\|^2] \leq 1 \ \ (\forall k
% \in {\bf N}_0),
%\end{align}
%%
%which implies
%%
%\begin{align}
%&
%E[\|x_c\|^2] \leq \|x_0\|^2/\delta \ \ (0\leq c <K).
%\end{align}
%%
%Hence, we have
%%%
%\begin{align}
%E[\|x_k\|^2]
%&<(2/\delta)\|x_{0}\|^2(2^{-1/K})^k \ \ (\forall k
% \in {\bf N}_0).
%\label{eq:proof-th1-expo-fin}
%\end{align}
%%
%This implies the 
%existence of $a=2/\delta$ and $\lambda=2^{-1/K}$ such
% that $a>0$, $0<\lambda<1$ and (\ref{eq:exp-def}) hold.
%Hence, by definition, the system is exponentially stable in the second moment.
%This completes the proof.
\end{IEEEproof}

Note that the above proof actually showed that the system is
exponentially stable if and only if (\ref{eq:asy-def}) holds; in other
words, (\ref{eq:stab-def}) was not used in the part ``1$\Rightarrow$2''.
This readily leads us to the following corollary as an
implicit result about asymptotic stability.

\begin{corollary}
\label{cr:asym}
Suppose $\xi$ satisfies Assumption~\ref{as:iid} and $A(\xi_k)$ satisfies
 Assumption~\ref{as:bound}.
The 
system (\ref{eq:fr-sys}) is
 asymptotically stable
 in the second moment if and only if (\ref{eq:asy-def}) holds.
\end{corollary}

The property described with (\ref{eq:asy-def}) is called attractivity \cite{Vidyasagar-book}.
Although asymptotic stability is usually defined
not only with attractivity but also with stability (as in
Definition~\ref{df:asym}),
it is known
in the deterministic systems case
that asymptotic stability can be ensured only with attractivity if the
system is linear and time-invariant.
Hence, the above corollary corresponds to a stochastic counterpart of
this conventional result because of Assumption~\ref{as:iid} (although system (\ref{eq:fr-sys}) itself is not time-invariant).

\subsection{Equivalence between Exponential Stability and Quadratic
  Stability}

The remaining issue in this section is to show the following theorem.

\begin{theorem}
\label{th:eqv-expo-quad}
Suppose $\xi$ satisfies Assumption~\ref{as:iid} and $A(\xi_k)$ satisfies
 Assumption~\ref{as:bound}.
The following two conditions are equivalent.
\begin{enumerate}
 \item 
The system (\ref{eq:fr-sys}) is exponentially stable in the second moment.
\item
The system (\ref{eq:fr-sys}) is quadratically stable.
\end{enumerate}
\end{theorem}

\begin{IEEEproof}
2$\Rightarrow$1:
A recursive use of (\ref{eq:quad-def})
leads to
\begin{align}
&
E[x_k^T P x_k] \leq \lambda^{2k}x_0^T P x_0\ \ (\forall k\in {\bf N}_0,
\forall x_0\in {\bf R}^n).
\end{align}
For the left-hand side of this inequality,
\begin{align}
&
\sigma_{\min}(P)E[\|x_k\|^2]\leq E[x_k^T P x_k],
\end{align}
while for the right-hand side,
\begin{align}
&
\lambda^{2k}x_0^T P x_0 \leq \sigma_{\max}(P)\|x_0\|^2 \lambda^{2k}.
\end{align}
Hence, we have (\ref{eq:exp-def}) with
$a=\sqrt{\sigma_{\max}(P)/\sigma_{\min}(P)}$ and the same $\lambda$,
which means by definition that
 the system is exponentially stable in the second moment.

\medskip
1$\Rightarrow$2:
Take a positive $\epsilon$ such that
$\lambda_\epsilon:=\lambda+\epsilon<1$ and define
\begin{align}
&
\Gamma^{k_2}_{k_1}:=
\begin{cases}
I & (k_2=k_1-1)\\
(A(\xi_{k_2})/\lambda_\epsilon)\ldots 
(A(\xi_{k_1})/\lambda_\epsilon)
& (k_2\geq k_1)
\end{cases}
\end{align}
for non-negative integers $k_1$ and $k_2\ (\geq k_1-1)$.
Then, (\ref{eq:exp-def}) can be rewritten as
\begin{align}
&
x_0^T E[(\Gamma^{k-1}_{0})^T \Gamma^{k-1}_{0}]x_0 \leq x_0^T
(a^2(\lambda^2/\lambda^2_\epsilon)^k I)x_0\notag \\ 
&
(\forall k\in {\bf N}_0,
\forall x_0\in {\bf R}^n),
\label{eq:th2-pf-rw}
\end{align}
where the well-definedness of the expectation in the left-hand side is ensured under
 Assumptions~\ref{as:iid} and \ref{as:bound} (in essentially the same manner as
Subsection~\ref{ssc:stab}).
Since $\xi$ satisfies Assumption~\ref{as:iid}, the above inequality leads to
\begin{align}
&
E[(\Gamma^{k_2}_{k_1})^T \Gamma^{k_2}_{k_1}] \leq
a^2(\lambda^2/\lambda^2_\epsilon)^{k_2-k_1+1} I\notag \\ 
&
(\forall k_1, k_2\in {\bf N}_0\ {\rm s.t.}\ k_2\geq k_1).
\label{eq:th2-pf-rw-any}
\end{align}

We next define
\begin{align}
&
P^K_k
:=
\lambda_\epsilon^{-2}I+
\lambda_\epsilon^{-2}(\Gamma^{k}_{k})^T \Gamma^{k}_{k}+
\ldots
+
\lambda_\epsilon^{-2}(\Gamma^{K}_{k})^T \Gamma^{K}_{k}
\label{eq:def-PKk}
\end{align}
for $k$ and $K\in {\bf N}_0$ such that $K\geq k\geq 0$.  Then, it satisfies
\begin{align}
&
\lambda^2_\epsilon P^K_k - A(\xi_k)^T P^K_{k+1} A(\xi_k) =I\notag \\ 
&
(\forall k, K\in {\bf N}_0\ {\rm s.t.}\ K>k\geq 0),
\end{align}
and it follows from (\ref{eq:fr-sys}) that
\begin{align}
&
\lambda^2_\epsilon E[x_{k} E[P^K_k] x_{k}] - E[x_{k+1}^T E[P^K_{k+1}]
x_{k+1}^T] \geq 0.
\label{eq:th2-pf-quad-K}
\end{align}
On the other hand,
(\ref{eq:def-PKk}) also implies that
the sequence of
\begin{align}
&
E[P^K_k]
=
\lambda_\epsilon^{-2}I+
\lambda_\epsilon^{-2}E[(\Gamma^{k}_{k})^T \Gamma^{k}_{k}]+
\ldots
+
\lambda_\epsilon^{-2}E[(\Gamma^{K}_{k})^T \Gamma^{K}_{k}]
\end{align}
with respect to $K\ (\geq k)$ for each fixed $k$ is
monotonically non-decreasing under the
semi-order relation based on positive semidefiniteness
(i.e., $E[P^K_k]\leq E[P^{K+1}_k]$).
In addition, 
it follows from
(\ref{eq:th2-pf-rw-any}) (and $a\geq 1$) that
\begin{align}
&
E[P^K_k]
\leq
\lambda_\epsilon^{-2}
a^2
(1+(\lambda^2/\lambda^2_\epsilon)+\ldots+(\lambda^2/\lambda^2_\epsilon)^{K-k+1})I,
\end{align}
whose right-hand side converges to a constant matrix as $K\rightarrow\infty$.
Hence, this sequence
also converges to a constant matrix as
$K\rightarrow\infty$.
Since this constant matrix does not depend on $k$
because of Assumption~\ref{as:iid}, we 
denote it by $P$, which is
obviously positive definite.
Then, 
letting $K\rightarrow \infty$ in
(\ref{eq:th2-pf-quad-K}) leads to
\begin{align}
&
\lambda^2_\epsilon E[x_{k} P x_{k}] - E[x_{k+1}^T P x_{k+1}^T] \geq 0,
\end{align}
which holds for every $k\in {\bf N}_0$.
Hence,
(\ref{eq:quad-def}) with $\lambda$ replaced by
$\lambda_\epsilon\ (<1)$ is satisfied,
which means by definition
that the system is quadratically stable.
This completes the proof.
\end{IEEEproof}

As stated at the beginning of this section,
no equivalence similar to that
in the above theorem holds
in the case with the usual deterministic linear time-varying systems.
Hence, this equivalence cannot be obtained without dealing
with randomness behind our system and thus the relevant stability
definitions for the system viewed as stochastic systems appropriately.
In particular, Assumption~\ref{as:iid} played a crucial role in showing
such equivalence.
To see this, let us temporarily consider a Markov chain $\xi$ (which fails to satisfy
 Assumption~\ref{as:iid}) and the associated system (\ref{eq:fr-sys}),
 which can be seen as the so-called Markov jump linear system \cite{Costa-book}.
Then, as is well known, the necessary and sufficient condition for
 exponential stability of the system can be described only with the
 mode-dependent Lyapunov matrix; this is true even when the Markov chain behind the
 system is time-homogeneous (i.e., stationary) and ergodic.
Hence, the quadratic stability defined with a constant Lyapunov matrix
 cannot be equivalent to exponential stability in such a case.
This in turn implies that assuming $\xi$ is stationary and ergodic is
insufficient for showing the
equivalence between quadratic stability and exponential stability,
and thus, Assumption~\ref{as:iid} is indeed essential.
In addition, it is also noted that deterministic LTI systems can be seen as a
special case of our systems with $\xi$ satisfying
Assumptions~\ref{as:iid} and \ref{as:bound} if we restrict our attention
to the distribution of $\xi_k$ that can take only a single value.
Since the definitions of stability in this paper immediately reduce to those for
deterministic LTI systems in that case, and since their equivalence is known to
hold, our results can be seen as a stochastic extension of such
conventional results.

\section{Stability Analysis Based on Lyapunov Inequality}
\label{sc:lyap}

Theorems~\ref{th:eqv-asym-expo} and \ref{th:eqv-expo-quad}
in the preceding section
showed the complete equivalence of the three stability
notions defined in Section~\ref{sc:sys} for system (\ref{eq:fr-sys})
under Assumptions~\ref{as:iid} and \ref{as:bound}.
Since the definition of quadratic stability is, unlike the other two,
expected to be compatible with 
the analysis based on Lyapunov inequalities,
we deal with this stability notion and discuss the corresponding Lyapunov
inequality in this section.

\subsection{Lyapunov Inequality for Quadratic Stability}

We first show the following theorem, which gives key inequality
conditions for stability analysis.

\begin{theorem}
\label{th:lyap}
Suppose $\xi$ satisfies Assumption~\ref{as:iid} and $A(\xi_k)$
satisfies Assumption~\ref{as:bound}.
The following three conditions are equivalent.
\begin{enumerate}
 \item 
The system (\ref{eq:fr-sys}) is quadratically stable.
\item
There exist $P\in {\bf S}^{n\times n}_+$ and $\lambda\in (0,1)$ such
 that
\begin{align}
&
E[\lambda^2 P - A(\xi_0)^T P A(\xi_0)]\geq 0.\label{eq:lyap-lambda}
\end{align}
\item
There exists $P\in {\bf S}^{n\times n}_+$ such
 that
\begin{align}
&
E[P - A(\xi_0)^T P A(\xi_0)]> 0.\label{eq:lyap}
\end{align}
\end{enumerate}
\end{theorem}

\begin{IEEEproof}
1$\Rightarrow$2:
Taking $k=0$ in inequality (\ref{eq:quad-def}) implies
\begin{align}
x_0^T E[\lambda^2 P - A(\xi_0)^T P A(\xi_0)] x_0 \geq 0\ \ (\forall
 x_0\in {\bf R}^n),
\end{align}
which is nothing but (\ref{eq:lyap-lambda}).

\medskip
2$\Rightarrow$1:
Since $\xi$ satisfies Assumption~\ref{as:iid},
(\ref{eq:lyap-lambda}) implies
\begin{align}
&
E[\lambda^2 P - A(\xi_k)^T P A(\xi_k)]\geq 0 \ \ \ (\forall k \in {\bf
 N}_0).
\label{eq:lyapunov-exp-allk}
\end{align}
Since $x_k$ and $A(\xi_k)$ are independent of each other, this further
 implies
\begin{align}
&
E[x_k^T(\lambda^2 P - A(\xi_k)^T P A(\xi_k))x_k]\geq 0 \ \ \ (\forall k \in {\bf
 N}_0),
\end{align}
which is nothing but (\ref{eq:quad-def}).

\medskip
2$\Leftrightarrow$3:
Adding $(1-\lambda^2)P>0$ to (\ref{eq:lyap-lambda}) immediately
leads to (\ref{eq:lyap}).  The opposite assertion is obvious.
\end{IEEEproof}

If $A(\xi_0)$ is deterministic, then (\ref{eq:lyap}) obviously
reduces to the usual Lyapunov inequality for deterministic linear
systems.
Hence, (\ref{eq:lyap}) is a natural extension of the usual Lyapunov
inequality for the stochastic systems case.
The well-definedness of the expectation in the Lyapunov inequality
(\ref{eq:lyap}) is
ensured under Assumption~\ref{as:bound}.

In addition, 
the proof on the equivalence between 1) and 2)
of the above theorem implies that, for each $\lambda$ and every $P>0$,
(\ref{eq:lyap-lambda}) holds if and only if (\ref{eq:quad-def})
holds.
This implies that the decay rate of 
the Lyapunov function in the definition of quadratic stability can be
evaluated in a necessary and sufficient sense through
(\ref{eq:lyap-lambda}).
This, together with the proof of Theorem~\ref{th:eqv-expo-quad}, further implies that
we can also evaluate the convergence rate of the sequence
$(\sqrt{E[\|x_k\|^2]})_{k\in {\bf N}_0}$ (i.e., minimal $\lambda$ satisfying
(\ref{eq:exp-def})) through (\ref{eq:lyap-lambda}).
Hence, the alternative representation (\ref{eq:lyap-lambda}) of the
Lyapunov inequality is also useful.

\subsection{Connections to Relevant Results}

Since our system description covers a wide class of discrete-time linear
systems with stochastic dynamics, the associated results can be seen as
a generalization of some existing results.
For instance, the following cases are relevant to our study.

{\it Case of Systems with State-Multiplicative Noise}:
Let us consider the $Z$-dimensional stochastic process $\xi$ 
satisfying Assumption~\ref{as:iid} and
\begin{align}
& E[\xi_0]=0,\ \ E[\xi_0 \xi_0^T]={\rm diag}(v_1,\ldots,v_Z),
\end{align}
where $v_i \in {\bf R}_+\ (i=1,\ldots,Z)$ are given constants.
For $\xi_k=[\xi_{1k},\ldots,\xi_{Zk}]^T$,
let us further consider the system (\ref{eq:fr-sys}) with
\begin{align}
&
A(\xi_k)=A_0 + \sum_{i=1}^Z A_i \xi_{ik},
\end{align}
where $A_i\in {\bf R}^{n\times n}\ (i=0,\ldots,Z)$ are given constant matrices.
This class of stochastic systems are called systems with
state-multiplicative noise; obviously, this class is a special case of
our systems.
Hence, it readily follows from Theorem~\ref{th:lyap} that the
system is quadratically (i.e., exponentially) stable if and only if
there exists $P\in {\bf S}^{n\times n}_+$ such that
\begin{align}
&
P-A_0^TPA_0-\sum_{i=1}^Z v_i A_i^T P A_i>0.
\end{align}
This LMI condition is nothing but that in Chapter~9 of \cite{Boyd-book}.

{\it Case of Switched Systems with i.i.d.\ Switching Signal}:
Let us next consider the $1$-dimensional stochastic process $\xi$ 
satisfying Assumption~\ref{as:iid} and
\begin{align}
&
\xi_k \sim D(d,p),\ d=[1,2,\ldots,S],\ p=[p_1,p_2,\ldots,p_S],
\end{align}
where $D(d,p)$ denotes the discrete distribution such that the
event $\xi_k=i$ occurs with probability $p_i$ for each $i=1,\ldots,S$.
Let us further consider the system (\ref{eq:fr-sys}) with
\begin{align}
&
A(\xi_k)= A_{[\xi_k]},
\end{align}
where $A_{[i]}\in {\bf R}^{n\times n}\ (i=1,\ldots,S)$ are given
constant matrices.
We see that the value $A_{[\xi_k]}$ is switched in accordance with the
i.i.d.\ switching signal $\xi$, and hence, the above system is a
switched system with an i.i.d.\ switching signal.
Since this system is also a special case of our systems, we can see that
the system is quadratically stable if and only if
there exists $P\in {\bf S}^{n\times n}_+$ such that
\begin{align}
&
P-\sum_{i=1}^S p_i A_{[i]}^T P A_{[i]}>0.
\end{align}
This LMI condition is nothing but that in Chapter~3 of \cite{Costa-book} (see Corollary~3.26).

\subsection{LMI Optimization}
\label{ssc:lmi-op}

We next discuss how to solve the Lyapunov inequality (\ref{eq:lyap-lambda}) or (\ref{eq:lyap})
for stability analysis of system (\ref{eq:fr-sys}).
As in the preceding subsection, our Lyapunov inequality
readily reduces to standard LMIs with
given deterministic (scalars and) matrices in cases with some specific systems.
In the general case, however,
the form of
inequalities (\ref{eq:lyap-lambda}) and (\ref{eq:lyap}), in which
the decision variable $P$ is contained in the operation of
expectation, makes it nontrivial to solve them.
This issue can be resolved as follows.

Let us first define
\begin{align}
&
A_{\rm e}(\xi_0):={\rm row}(A(\xi_0))^T {\rm row}(A(\xi_0)),
\label{eq:equiv-rep-vecvec}
\end{align}%
whose elements 
cover all the second degree products of the elements of $A(\xi_0)$.
$E[A_{\rm e}(\xi_0)]$ is well-defined by
Lemma~\ref{lm:product-bound} under
Assumption~\ref{as:bound} and becomes a positive semidefinite matrix.
Let us further take $\bar{A}\ (\in{\bf R}^{n^2\times n^2})$ such that
\begin{align}
&
\bar{A}^T \bar{A}=E[A_{\rm e}(\xi_0)],
\label{eq:equiv-rep-decom}
\end{align}
and introduce the following partitioning of $\bar{A}$.
\begin{align}
&
\bar{A}=:
\left[\bar{A}_1, \bar{A}_2, \ldots, \bar{A}_n\right] \ \ 
(\bar{A}_i \in {\bf R}^{n^2\times n}\ (i=1,\ldots,n))
\end{align}
Then, for
\begin{align}
\bar{A}^\prime :=[\bar{A}_1^T, \bar{A}_2^T, \ldots,
 \bar{A}_n^T]^T \in {\bf R}^{n^3\times n},
\end{align}
the matrix
\begin{align}
&
(\bar{A}^\prime)^T
(P\otimes I_{n^2}) 
\bar{A}^\prime
\label{eq:equiv-rep1}
\end{align}
with the decision variable $P$ can be confirmed
to coincide with $E[A(\xi_0)^T P A(\xi_0)]$.
In addition, another representation of $E[A(\xi_0)^T P A(\xi_0)]$ can be
also given by
\begin{align}
&
\bar{A}_{\rm e2}(I_n\otimes {\rm row}(P)^T)
\label{eq:equiv-rep2}
\end{align}%
for $\bar{A}_{\rm e2}=E[A_{\rm e2}(\xi_0)]\in {\bf R}^{n\times n^3}$,
where
\begin{align}
&
A_{\rm e2}(\xi_0):=
\begin{bmatrix}
{\rm row}(a_1 a_1^T) & \cdots & {\rm row}(a_n a_1^T) \\
\vdots & \ddots & \vdots \\
{\rm row}(a_1 a_n^T) & \cdots & {\rm row}(a_n a_n^T)
\end{bmatrix}
\end{align}%
under the partitioning $A(\xi_0)=:[a_1, a_2, \ldots, a_n]$ ($\xi_0$
is omitted in the column random vectors for notation simplicity).

Although (\ref{eq:equiv-rep1}) has a form compatible with the extension
toward stabilization synthesis discussed in the next section, (\ref{eq:equiv-rep2})
has the advantage that we do not need to decompose matrices as in
(\ref{eq:equiv-rep-decom}).
The above arguments can be summarized by the following lemma.
\begin{lemma}
\label{lm:lyap-lmi}
For given $P$, the expectation
$E[A(\xi_0)^T P A(\xi_0)]$ is equivalent to
 (\ref{eq:equiv-rep1}) and (\ref{eq:equiv-rep2}).
\end{lemma}

The important point here is that in both (\ref{eq:equiv-rep1}) and
(\ref{eq:equiv-rep2}),
the decision variable $P$ has been taken out from the expectation operation.
The implication is that once we calculate 
$\bar{A}^\prime$ in (\ref{eq:equiv-rep1}) or
$\bar{A}_{\rm e2}$ in (\ref{eq:equiv-rep2}), we can
then solve (\ref{eq:lyap-lambda}) and (\ref{eq:lyap}) as the standard
linear matrix inequalities (LMIs).

\section{Stabilization State Feedback Synthesis Based on Lyapunov Inequality}
\label{sc:syn}

In this section, we discuss stabilization
state feedback synthesis based on the Lyapunov inequality condition
derived in the preceding section.

\subsection{Problem of Stabilization State Feedback Synthesis}

We first state the synthesis problem to be tackled in this section.
Let us consider the $Z$-dimensional process
$\xi$ satisfying Assumption~\ref{as:iid} and the associated system
\begin{align}
&
x_{k+1} = A_{\rm op}(\xi_k) x_k + B_{\rm op}(\xi_k) u_k,\label{eq:open-sys}
\end{align}
where $x_k \in {\bf R}^n$, $u_k \in {\bf R}^{m}$, 
$A_{\rm op}:{\boldsymbol {\mit\Xi}} \rightarrow {\bf R}^{n\times n}$ and
$B_{{\rm op}}:{\boldsymbol {\mit\Xi}} \rightarrow {\bf R}^{n\times m}$.
On the coefficient matrices of the above system, we make the following
assumption similar to Assumption~\ref{as:bound}.

\begin{assumption}
\label{as:inf-syn}
The squares of elements of 
$A_{\rm op}(\xi_k)$ and $B_{\rm op}(\xi_k)$ are all Lebesgue integrable.
\end{assumption}

Let us consider the state feedback 
\begin{align}
&
u_k=F x_k \label{eq:state-feedback}
\end{align}
with the static time-invariant gain $F\in {\bf R}^{m\times n}$.
The closed-loop system can be described by
(\ref{eq:fr-sys}) with
\begin{align}
&
A(\xi_k)=A_{\rm op}(\xi_k)+B_{\rm op}(\xi_k)F.
\label{eq:closed-loop}
\end{align}
Note that if $A_{\rm op}(\xi_k)$ and $B_{\rm op}(\xi_k)$ satisfy Assumption~\ref{as:inf-syn}
then the above $A(\xi_k)$ also satisfies Assumption~\ref{as:bound} (for
each fixed $F$) by Lemma~\ref{lm:product-bound}.
This section studies the synthesis problem of 
$F$ such that the closed-loop system is quadratically stable.

\subsection{LMI for Synthesis}

For a given $F\in {\bf R}^{m\times n}$, it readily follows from
Theorem~\ref{th:lyap} that
the closed-loop system is quadratically stable if and only if
there exists $P\in {\bf S}^{n\times n}_+$ such that
\begin{align}
&
E[P-(A_{\rm op}(\xi_0)+B_{\rm op}(\xi_0)F)^T P (A_{\rm op}(\xi_0)+B_{\rm
op}(\xi_0)F)]\!>\!0.
\label{eq:lyap-syn}
\end{align}
Hence, our synthesis problem reduces to that of
searching for $F$ such that there exists $P>0$ satisfying the
above inequality.
Since the inequality not only involves the expectation operation but also is
nonlinear in the decision variables $P$ and $F$, 
it is more difficult to deal with than (\ref{eq:lyap}) about the
analysis.
Fortunately, however, a technique similar to (\ref{eq:equiv-rep1}) can
indeed lead us to an alternative representation of (\ref{eq:lyap-syn})
that is compatible with the Schur complement technique
\cite{Boyd-book}.

To see this, let us first define
\begin{align}
G_{\rm e}(\xi_0)
:=&
[{\rm row}(A_{\rm op}(\xi_0)), {\rm row}(B_{\rm op}(\xi_0))]^T
\notag \\
&\cdot [{\rm row}(A_{\rm op}(\xi_0)), {\rm row}(B_{\rm op}(\xi_0))],
\label{eq:equiv-rep-vecvec-syn}
\end{align}%
whose elements cover all the second order products of the elements of
$[A_{\rm op}(\xi_0), B_{\rm op}(\xi_0)]$.
$E[G_{\rm e}(\xi_0)]$ is well-defined under
Assumption~\ref{as:inf-syn} and becomes a positive semidefinite matrix.
Let us further take $\bar{G}\ (\in {\bf R}^{(n+m)n\times(n+m)n})$ such
that
\begin{align}
&
\bar{G}^T \bar{G}=E[G_{\rm e}(\xi_0)],
\label{eq:equiv-rep-decom-syn}
\end{align}
and introduce the following partitioning of $\bar{G}$.
\begin{align}
&
\bar{G}=:
\left[\bar{G}_{A1}, \ldots, \bar{G}_{An}, \bar{G}_{B1}, \ldots,
\bar{G}_{Bn}\right] \notag\\
& 
(\bar{G}_{Ai} \in {\bf R}^{(n+m)n\times n}, \bar{G}_{Bi} \in {\bf
R}^{(n+m)n\times m}\ (i=1,\ldots,n))
\end{align}
Then, for
\begin{align}
&
\bar{G}^\prime_A :=[\bar{G}_{A1}^T, \ldots,
 \bar{G}_{An}^T]^T \in {\bf R}^{(n+m)n^2 \times n},\\
&
\bar{G}^\prime_B :=[\bar{G}_{B1}^T, \ldots,
\bar{G}_{Bn}^T]^T \in {\bf R}^{(n+m)n^2 \times m},
\label{eq:def-GpB}
\end{align}
the matrix
\begin{align}
&
\left(\bar{G}^\prime_{A}+\bar{G}^\prime_{B} F\right)^T (P\otimes I_{(n+m)n})
\left(\bar{G}^\prime_{A}+\bar{G}^\prime_{B} F\right)
\label{eq:equiv-rep-syn}
\end{align}
with the decision variables $P$ and $F$ can be confirmed
to coincide with 
$E[(A_{\rm op}(\xi_0)+B_{\rm
op}(\xi_0)F)^T P (A_{\rm op}(\xi_0)+B_{\rm
op}(\xi_0)F)]$.
Hence, once we calculate $\bar{G}^\prime_{A}$ and
$\bar{G}^\prime_{B}$, 
the inequality condition
(\ref{eq:lyap-syn}) can be dealt with as a standard
matrix inequality; in particular, the resulting inequality has a
form compatible with the Schur complement technique.

Since $P\otimes I_{(n+m)n} >0$ for $P\in {\bf S}^{n\times n}_+$,
the above arguments lead us to the following lemma.
\begin{lemma}
\label{lm:extension}
For given $P\in {\bf S}^{n\times n}_+$ and
$F\in {\bf R}^{m\times n}$, (\ref{eq:lyap-syn}) holds
if and only if 
\begin{align}
&
\begin{bmatrix}
P& \ast\\
(P\otimes I_{(n+m)n})
\left(\bar{G}^\prime_{A}+\bar{G}^\prime_{B} F\right) &
P\otimes I_{(n+m)n}
\end{bmatrix}>0,
\label{eq:extension}
\end{align}
where $\ast$ denotes the transpose of the lower left block in the matrix.
\end{lemma}

This lemma, together with
the congruence transformation with ${\rm diag}(X, X\otimes I_{(n+m)n})$
for $X=P^{-1}$ and the change of variables $Y=FX$,
further leads us to the following theorem about the synthesis.
\begin{theorem}
\label{th:syn}
Suppose $\xi$ satisfies Assumption~\ref{as:iid} and $A_{\rm op}(\xi_k)$
and $B_{\rm op}(\xi_k)$
satisfy Assumption~\ref{as:inf-syn}.
There exists a gain $F$ such that the closed-loop system
(\ref{eq:fr-sys}) with (\ref{eq:closed-loop}) is quadratically stable
if and only if there exist $X\in {\bf S}^{n\times n}_+$ and
$Y\in {\bf R}^{m\times n}$ satisfying
\begin{align}
&
\begin{bmatrix}
X& \ast\\
\bar{G}^\prime_{A}X+\bar{G}^\prime_{B} Y &
X\otimes I_{(n+m)n}
\end{bmatrix}>0 \label{eq:lmi-syn}
\end{align}
for $\bar{G}^\prime_{A}$ and $\bar{G}^\prime_{B}$ defined by (\ref{eq:equiv-rep-vecvec-syn})--(\ref{eq:def-GpB}).
In particular, $F=YX^{-1}$ is one such stabilization gain.
\end{theorem}

Although the above theorem is derived from (\ref{eq:lyap}) without
$\lambda$, the same technique can be applied also to
(\ref{eq:lyap-lambda}) with $\lambda$, which leads to the following
corollary.
\begin{corollary}
\label{cr:syn}
Suppose $\xi$ satisfies Assumption~\ref{as:iid} and $A_{\rm op}(\xi_k)$
and $B_{\rm op}(\xi_k)$
satisfy Assumption~\ref{as:inf-syn}.
There exists a gain $F$ such that the corresponding closed-loop system
is quadratically stable if and only if
there exist $X\in {\bf S}^{n\times n}_+$,
$Y\in {\bf R}^{m\times n}$ 
and $\lambda\in (0,1)$
satisfying
\begin{align}
&
\begin{bmatrix}
\lambda^2 X& \ast\\
\bar{G}^\prime_{A}X+\bar{G}^\prime_{B} Y &
X\otimes I_{(n+m)n}
\end{bmatrix}\geq0.\label{eq:lmi-syn-lambda}
\end{align}
In particular, $F=YX^{-1}$ is one such stabilization gain.
\end{corollary}

If we aim not only at stabilizing the closed-loop system but also at
minimizing $\lambda$ in (\ref{eq:quad-def})
(which corresponds to the convergence rate related to the definition of exponential
stability by Theorem~\ref{th:eqv-expo-quad}), this corollary will play an important role.

\section{Numerical Examples}
\label{sc:example}

This section is devoted to numerical examples.
We first numerically demonstrate with a simple
example that the Lyapunov inequality (\ref{eq:lyap-lambda}) gives a necessary and
sufficient condition for quadratic stability (and thus exponential
stability) of system (\ref{eq:fr-sys})
as indicated by Theorems~\ref{th:eqv-expo-quad} and \ref{th:lyap}.
Then, we provide a more challenging example for motivating our study,
in which the stabilization state feedback is designed for the discrete-time system obtained through discretizing a
continuous-time deterministic linear system with a randomly time-varying
sampling interval.

\subsection{Demonstration of Strictness in Stability Analysis Based on
  Lyapunov Inequality}

Let us consider the $2$-dimensional stochastic process $\xi$ 
that satisfies Assumption~\ref{as:iid} and is given by the sequence of
$\xi_k=[\xi_{1k}, \xi_{2k}]^T,\
\xi_{1k}\sim N(\mu,\sigma^2)\ (\mu=0, \sigma=0.2),\
\xi_{2k}\sim U(\underline{d},\overline{d})\ (\underline{d}=-0.5, \overline{d}=0.5)$,
where $N(\mu,\sigma^2)$ and $U(\underline{d},\overline{d})$ respectively
denote the normal distribution with mean $\mu$
and standard deviation $\sigma$ and
the continuous uniform distribution with minimum $\underline{d}$ and
maximum $\overline{d}$.
Let us further consider the stochastic system (\ref{eq:fr-sys}) with
\begin{align}
&
A(\xi_k)=
\begin{bmatrix}
0.3+\xi_{2k} & 0.8+\xi_{1k} &    -0.5\\
0.5  &   0.3+\xi_{1k}\xi_{2k} & -1.2+(\xi_{1k})^2\\
-0.2 &   0.8 &    0.6
\end{bmatrix}.
\end{align}
Through numerical
stability analysis of this system, we discuss the strictness of
our Lyapunov inequality condition.

\begin{figure}[t]
  \centering
  \includegraphics[width=0.8\linewidth]{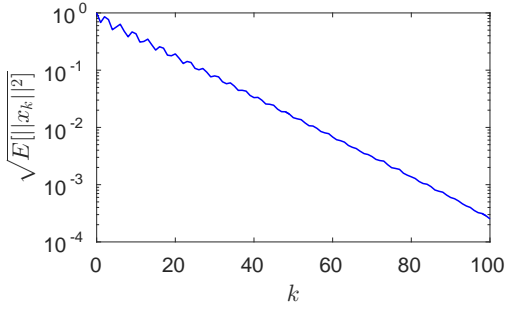}
  \caption{Time response of estimate of $\sqrt{E[||x_k||^2]}$ calculated with
 $10^5$ sample
 paths of $\xi$.}
  \label{fig:mscs17_si}
\end{figure}

We first search for the minimal $\lambda$ such that there exists $P>0$
satisfying (\ref{eq:lyap-lambda}).
As stated in Lemma~\ref{lm:lyap-lmi}, 
the matrix (\ref{eq:equiv-rep1}) is an alternative representation of
the expectation $E[A(\xi_0)^T P
A(\xi_0)]$ in (\ref{eq:lyap-lambda}).
Hence, once we calculate $\bar{A}^{\prime}$ in (\ref{eq:equiv-rep1})
for the above $A(\xi_0)$, it readily follows that we can 
solve (\ref{eq:lyap-lambda}) as an LMI for each fixed $\lambda$.
This enables us to achieve
the aforementioned minimization through a bisection method with
respect to $\lambda$; the resulting minimum is expected to
correspond to the convergence rate with respect to
$\left(\sqrt{E[||x_k||^2]}\right)_{k \in {\bf N}_0}$
 by Theorems~\ref{th:eqv-expo-quad} and
\ref{th:lyap}.
We performed calculation of $\bar{A}^{\prime}$ with MATLAB and 
Symbolic Math Toolbox,
and minimized $\lambda$ with MATLAB,
YALMIP \cite{YALMIP} and SDPT3 \cite{SDPT3}.
Then, the minimal $\lambda$ became 0.9219$\ (<1)$,
which implies
exponential stability of the system
 by Theorems~\ref{th:eqv-expo-quad} and
\ref{th:lyap}.

We next confirm that the above minimal $\lambda$ indeed corresponds to
the convergence rate with respect to
$\left(\sqrt{E[||x_k||^2]}\right)_{k \in {\bf N}_0}$, through
calculating its estimate $\lambda_{\rm est}$ with $N_{\rm s}$ sample paths of $\xi$.
For $N_{\rm s}=10^5$, such sample-based estimation
of $\sqrt{E[||x_k||^2]}$ provided us with the time response shown in
Fig.~\ref{fig:mscs17_si},
where we took $x_0=[1,0,0]^T$ as the initial state of the system.
As we can see from the figure, 
the estimate of 
$\sqrt{E[||x_k||^2]}$ decays with an almost constant rate
after the elapse of sufficient time.
The decay rate $\lambda_{\rm est}$ obtained from the data at $k=50$ and $100$ in 
Fig.~\ref{fig:mscs17_si} was 0.9213
(similar results were obtained
regardless of the initial state under the same sample of $\xi$).
Since this value is close to the above minimization result,
it numerically suggests that
the minimal $\lambda$ obtained with the Lyapunov
inequality corresponds to the true convergence rate.

\subsection{Stabilization of Discrete-Time System Obtained under
  Randomly Time-Varying Sampling Interval}

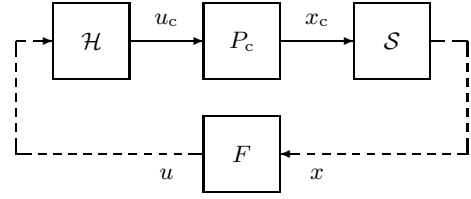
\begin{figure}[t]
  %\begin{minipage}{0.5\hsize}
\setlength{\unitlength}{1cm}
	\centering
	\scalebox{1}{\begin{picture}(6,2.5)(0,0)
\put(0.5,1.5){\framebox(1,1){${\cal H}$}}
\put(1.5,2){\vector(1,0){1}}
\put(2,2.2){\makebox(0,0)[b]{$u_{\rm c}$}}
\put(2.5,1.5){\framebox(1,1){$P_{\rm c}$}}
\put(3.5,2){\vector(1,0){1}}
\put(4,2.2){\makebox(0,0)[b]{$x_{\rm c}$}}
\put(4.5,1.5){\framebox(1,1){${\cal S}$}}
\put(5.5,2){\line(1,0){0.1}} \multiput(5.6,2)(0.2,0){1}{\line(1,0){0.1}} \put(5.8,2){\line(1,0){0.2}}
\put(6,2){\line(0,-1){0.1}} \multiput(6,1.9)(0,-0.2){6}{\line(0,-1){0.1}} \put(6,0.7){\line(0,-1){0.2}}
\put(6,0.5){\line(-1,0){0.03}} \multiput(5.97,0.5)(-0.2,0){11}{\line(-1,0){0.1}} \put(3.77,0.5){\vector(-1,0){0.27}}
\put(4,0.3){\makebox(0,0)[t]{$x$}}
\put(2.5,0){\framebox(1,1){$F$}}
\put(2.5,0.5){\line(-1,0){0.1}} \multiput(2.4,0.5)(-0.2,0){11}{\line(-1,0){0.1}} \put(0.2,0.5){\line(-1,0){0.2}}
\put(2,0.3){\makebox(0,0)[t]{$u$}}
\put(0,0.5){\line(0,1){0.1}} \multiput(0,0.6)(0,0.2){6}{\line(0,1){0.1}} \put(0,1.8){\line(0,1){0.2}}
\put(0,2){\line(1,0){0.03}} \multiput(0.03,2)(0.2,0){1}{\line(1,0){0.1}} \put(0.23,2){\vector(1,0){0.27}}
\end{picture}}  
	\caption{Sampled-data system with sampler and hold running under
 randomly time-varying sampling interval.}
	\label{fig:sdsystem}
  %\end{minipage}
\end{figure}
\begin{figure}[t]
  \centering
  \includegraphics[width=0.85\linewidth]{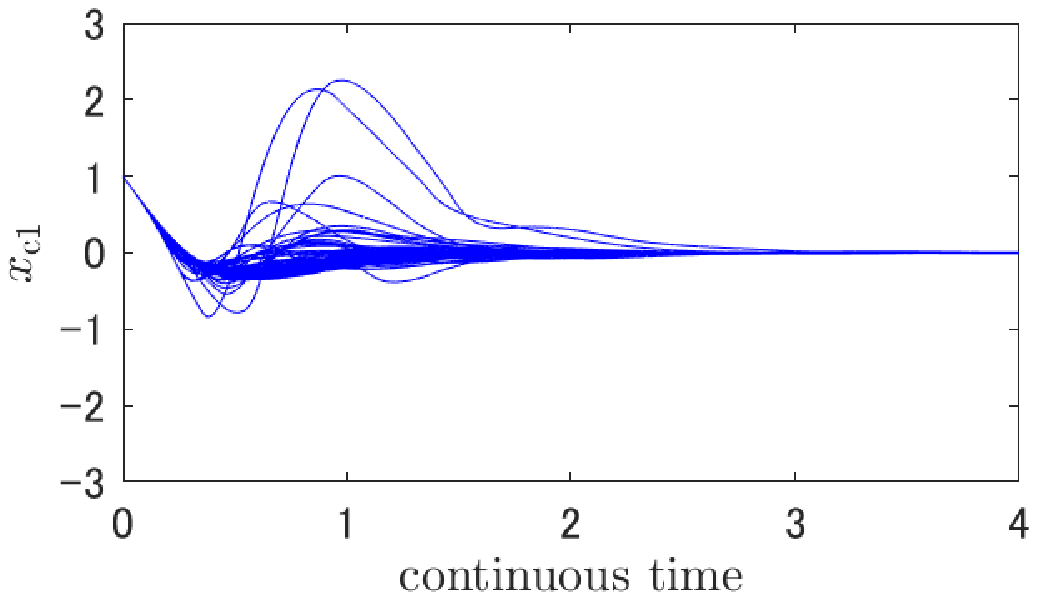}
  \includegraphics[width=0.85\linewidth]{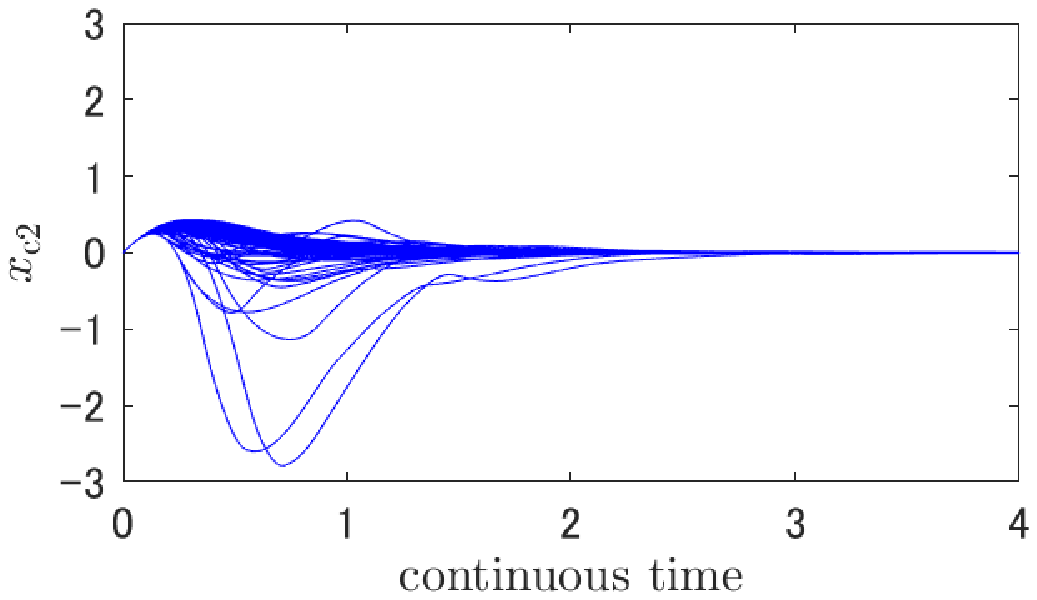}
  \includegraphics[width=0.85\linewidth]{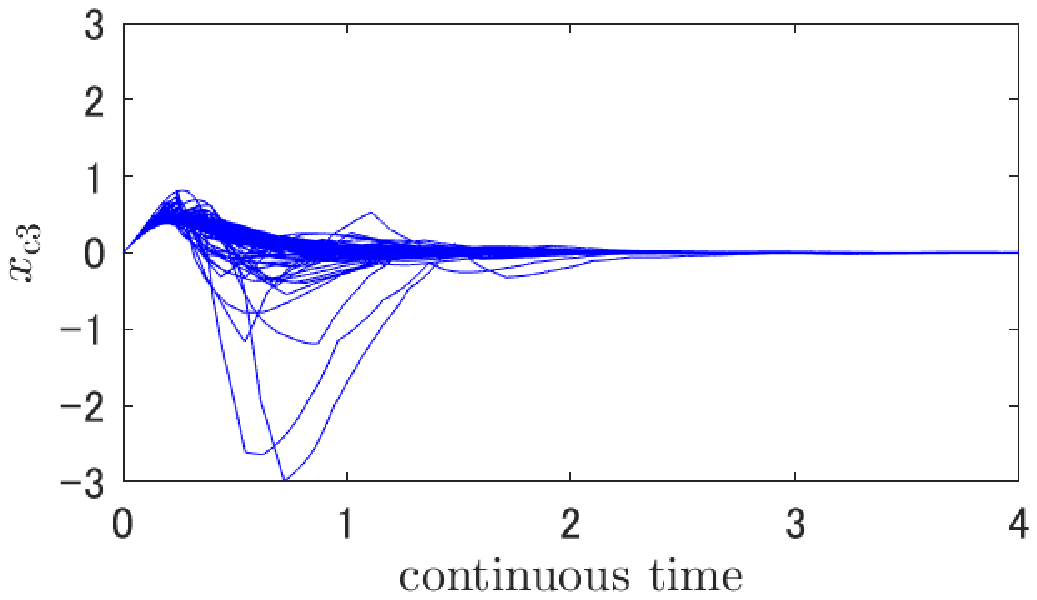}
  \includegraphics[width=0.85\linewidth]{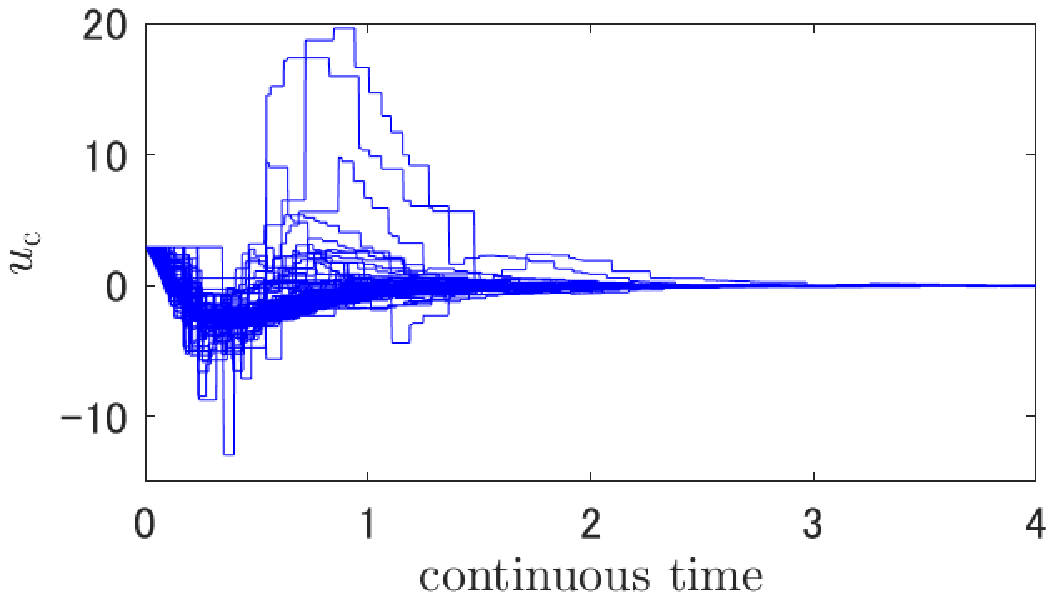}
  \caption{Overlays of responses of continuous-time
 $x_{\rm c}$ and $u_{\rm c}$ 
generated with 100 sample paths of $\xi$
and the initial state $x_{\rm c}(t_0)=[1,0,0]^T$.}
  \label{fig:sdsim}
\end{figure}

Let us next consider the sampled-data system, shown in Fig.~\ref{fig:sdsystem},
consisting of
the continuous-time deterministic linear unstable system 
$P_{\rm c}$ given by
\begin{align}
&
\dot{x}_{\rm c}=A_{\rm c}x_{\rm c} + B_{\rm c}u_{\rm c},\ \ 
A_{\rm c}=
\begin{bmatrix}
-4 & 3 & -8 \\
3 & 7 & -6 \\
0 & 8 & -2
\end{bmatrix},\ \ 
B_{\rm c}=
\begin{bmatrix}
0 \\ 0 \\ 1
\end{bmatrix},
\end{align}
the static time-invariant 
state feedback gain $F$ to be designed, 
and the sampler ${\cal S}$ and the zero-order hold ${\cal H}$ running
with the sampling instants $t_k\ (k\in {\bf N}_0)$, where
\begin{align}
&
t_0=0,\ \ t_{k+1}-t_k>0,\ \ \lim_{k\rightarrow \infty}t_k=\infty.
\end{align}
The relation between the continuous-time signals and
the discrete-time signals in Fig.~\ref{fig:sdsystem} is described as follows.
\begin{align}
&
x_k=x_{\rm c}(t_k),\ \ 
u_{\rm c}(t)=u_k\ \ 
(t\in [t_k, t_{k+1}); k\in {\bf N}_0)
\end{align}
For such a sampled-data system, we assume that
the sampling interval $h_k=t_{k+1}-t_k$ is randomly time-varying
(i.e., the
random case of aperiodic sampling
\cite{Montestruque-TAC04,Hetel-Auto17}) 
and given by
$h_k=h(\xi_k)=0.01+\xi_k$ with the $1$-dimensional stochastic process
$\xi$ that satisfies Assumption~\ref{as:iid} and
$\xi_k \sim {\rm Exp}(\nu)\ (\nu=20)$, where 
${\rm Exp}(\nu)$ denotes the exponential distribution with expectation $1/\nu$.
In this subsection, we consider designing $F$ stabilizing this
sampled-data stochastic system;
if we focus only on the signal values at the sampling
instants, this synthesis problem reduces to that of designing a state
feedback (\ref{eq:state-feedback}) stabilizing the discrete-time stochastic system (\ref{eq:open-sys}) with the random coefficients
\begin{align}
&
A_{\rm op}(\xi_k)=
e^{A_{\rm c}h(\xi_k)},\ \ 
B_{\rm op}(\xi_k)=
\int_0^{h(\xi_k)}e^{A_{\rm c}t}B_{\rm c}dt.
\end{align}

For the above discrete-time system, we searched for 
a solution of (\ref{eq:lmi-syn-lambda}) minimizing $\lambda$
with MATLAB, Symbolic Math Toolbox, YALMIP and SDPT3,
where matrix exponentials were dealt with through the second-order 
Pad\'{e} approximation in the computation.
Then, we obtained the gain
$F=\left[2.9242, 4.9123, -10.0501\right]$
with $\lambda=0.9193$, which implies the stability of the corresponding
discrete-time closed-loop system by Corollary~\ref{cr:syn}.
Since our control approach is developed for discrete-time stochastic
systems,
it can only ensure the convergence of the state
of the sampled-data system (in the stochastic sense) with respect to the sampling instants immediately.
Fortunately, however, the responses of the continuous-time signals $x_{\rm c}$
and $u_{\rm c}$ in the present sampled-data
system (with the above $F$) indeed converged to zero in the simulations of the authors;
Fig.~\ref{fig:sdsim} shows the overlays of
the responses of $x_{\rm c}$ and $u_{\rm c}$
generated with 100 sample paths of $\xi$
and the initial state $x_{\rm c}(t_0)=[1,0,0]^T$.
Since there is virtually no limitation on the class of continuous-time
linear systems (and that of
the distributions of $h_k$) in the above synthesis,
other synthesis problems could also be solved in a similar fashion.
This suggests strong potential of the proposed approach.

\section{Conclusion}

In this paper, we first showed that asymptotic stability, exponential
stability and quadratic stability are all equivalent for
discrete-time linear systems with stochastic dynamics
under the assumption that the underlying process $\xi$ has an
i.i.d.\ property.
Then, we discussed a Lyapunov inequality that can characterize
stability in a necessary and sufficient sense.
Our Lyapunov inequality
readily reduces to standard LMIs for some subclasses of stochastic systems.
In the general case, however,
the original form of the inequality seemed unsuitable
for numerical stability analysis because it must be solved for decision variables
contained in the expectation operation.
Hence, we also provided an idea to solve the inequality as a standard LMI
even in the general case; this idea was also used in the extension of
the Lyapunov inequality condition toward stabilization state feedback
synthesis.

\ifCLASSOPTIONcaptionsoff
  \newpage
\fi

\vfill

% Can be used to pull up biographies so that the bottom of the last one
% is flush with the other column.
%\enlargethispage{-5in}

% that's all folks
\end{document}